\documentclass[aps,pre,groupedaddress,article]{revtex4}
\usepackage{amsmath}
\usepackage{graphicx}
\usepackage{dcolumn}
\usepackage{bm}
\begin{document}

\title{Effects of kinked linear defects on planar flux line arrays}
\author{Eleni Katifori} \email{katifori@fas.harvard.edu}
\author{David R. Nelson}
\affiliation{Department of Physics, Harvard University, Cambridge, Massachusetts, 02138}

\begin{abstract}
In the hard core limit, interacting vortices in planar type II superconductors can be modeled as non-interacting one dimensional fermions propagating in imaginary time. 
We use this analogy to derive analytical expressions for the probability density and imaginary current of vortex lines interacting with an isolated bent line defect and to understand the pinning properties of such systems. 
When there is an abrupt change of the direction of the pinning defect, we find a sinusoidal modulation of the vortex density in directions both parallel and perpendicular to the defect.      
\end{abstract}

\maketitle

\section{Introduction}
\label{sec:Introduction}

The statistical physics of flux lines in high Tc superconductors has attracted considerable experimental and theoretical attention. Controlling the pinning properties of the magnetic vortices that penetrate the superconducting material above a critical external field $H_{c1}$ can be the key in achieving dissipationless electrical current flow~\cite{Blatter94}.  
The interplay between point and correlated disorder, thermal fluctuations and vortex-vortex repulsion controls the flux line configurations and leads to a variety of different phases~\cite{Fish91,Hwa93,Nelson93}.

Recent advances in manufacturing of high quality thin superconducting slabs and films, and experimental techniques for probing such systems, has made possible to observe mesoscopic vortex dynamics and image individual vortices confined to two dimensions~\cite{Bolle99,Tonomura01}.
The trajectories of vortices in these 2-d systems can be mapped to the world lines of 1-d bosons and in that sense planar superconductors can be an interesting laboratory for Luttinger liquid physics. 
These 1+1 dimensional systems have been extensively studied in the presence of correlated and point disorder~\cite{Polkovnikov05,Refael06,Hwa93b,Derev94}, or an isolated straight columnar pin ~\cite{Hofstetter04,Affleck04,Radzihovsky06}.

The continuum elastic energy for the coarse-grained displacement field $u(x,\tau)$ for such a system of vortex lines in the presence of a straight columnar pin at $x=0$ and a transverse magnetic field $H_{\perp}$ is
\begin{equation}\label{ContFreeEn}
F=\int \mathrm{d}x\mathrm{d}\tau \left[\frac{c_{44}}{2}(\partial_{\tau}u)^2+\frac{c_{11}}{2}(\partial_{x}u)^2-\frac{n_0 \phi_0 H_{\perp}}{4\pi}
(\partial_{\tau}u)\right]
-\epsilon_d n_0\int\mathrm{d}\tau \cos[2\pi n_0 u(0,\tau)]
\end{equation}
where $c_{11}$ and $c_{44}$ are the compressional and tilt moduli, $n_0$ is the average vortex density which depends linearly on
the external magnetic field, $\phi_0$ is the quantum flux and $\epsilon_d$ is the pinning strength.

A Luttinger liquid parameter g 
\begin{equation}\label{LuttPar}
g=\frac{\pi T n_0^2}{\sqrt{c_{11}c_{44}}}
\end{equation}
can be defined from this long distance free energy. The long distance probability density derived from (\ref{ContFreeEn}) exhibits Friedel-like oscillations, modulated by an exponential:
\begin{equation}\label{FriedelOsc}
\langle n(x)\rangle-n_0\sim \frac{\cos(2\pi n_0 x)}{|x|^{\alpha}}\exp(-|x|/\xi).
\end{equation}
The exponent $\alpha$ assumes different values for values of $g$ above and below unity: $\alpha=2g-1$ for $g<1$ and 
$\alpha=g$ for $g>1$. Here, $\xi$ is a coherence length inversely proportional to the transverse magnetic field, $\xi\sim H_{\perp}$ or equivalently 
the relative tilt between the columnar pin and the direction $\vec{H}$ of the in-plane magnetic field of the slab.

When the vortex array is dilute enough, or in the presence of a short range hard core repulsive interaction, the physics maps onto the free fermion problem. For a discussion in the context of commensurate-incommensurate transitions in adsorbed monolayers, see Refs~\cite{Pokrovsky79,Pokrovsky84,Coppersmith82,Schulz82}. In that case the compressional and tilt moduli are independent of the interaction potential details and assume values such that $g=1$.
The probability density distribution can be easily evaluated in this regime by use of the left and right eigenstates of the 
single particle non-Hermitian Hamiltonian
\begin{equation}\label{Hami}
\mathcal{H}(\tau)= -\frac{T^2}{2\gamma}\:\frac{\partial^2}{\partial x^2}-h(\tau)\:T\:\frac{\partial}{\partial x}-V_o\:\delta(x).
\end{equation}
where we have set $k_B=1$, $T$ is the temperature and $\gamma$ is proportional to the line tension of the vortices.
The simplest situation arises when the relative tilt between the defect and $\vec{H}$ remains constant. This condition  applies to the straight defect case, i.e. a simple columnar pin, in which case $h(\tau)\propto H_{\perp}$ is independent of $\tau$.
However, nanolithographic techniques could allow for the controlled fabrication of meandering linear defects, which motivates us to study the effect of a sudden change of direction or an abrupt termination of a defect trajectory within the sample. Because the one-dimensional Luttinger liquid is a critical system (with $g$-dependent exponents), there can be a striking response to such perturbations.

The case of a single flux line interacting with a meandering linear defect has been studied in Ref.~\cite{Katifori06}. In this paper, we study the more experimentally relevant case of many flux lines interacting with a single kinked or terminating defect, confined to a thin superconducting slab. 
In Section \ref{sec:ProbDistr} we derive an analytic expression for the probability density distribution and imaginary current of flux lines in the presence of a bent pinning defect. 
These analytic expressions are used in Section \ref{sec:Oscill} to understand how the bent line defect perturbs the (1+1)-dimensional vortex configurations. In Sec.~\ref{sec:Experiments}, we discuss potential experiments and defect lines that terminate.

\section{probability distribution of fluctuating vortex filaments}
\label{sec:ProbDistr}

In a spirit similar to Ref.~\cite{Katifori06}, where the interaction of a single flux line with a meandering linear defect with trajectory $x_o(\tau)$ was discussed, in this Section we derive an expression for the probability density of an array of flux lines interacting with an attractive delta function potential whose direction changes suddenly along the superconducting slab. To allow for a simple analytical treatment, we consider only piece-wise constant defect trajectories, and focus in particular on a defect consisting of two straight segments joined at an angle. Unlike the single vortex system, in the many flux line case there is no critical tilt that defines the critical point of a delocalization transition. The large number of extended states that are occupied even in the ground state in the many particle system wash out the delocalization transition of the single bound state.

\begin{figure}
\includegraphics[scale=0.8]{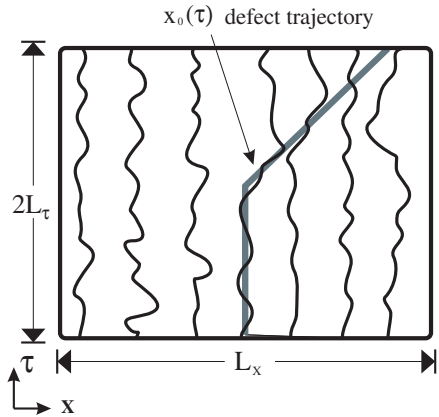}
\caption{\label{fig:ManyLines} Vortices confined to a planar superconductor. A linear defect (depicted with a thick grey line) that changes orientation in the superconducting slab acts as a pinning potential. The external magnetic field $H$ is along the $\tau$ dimension, which corresponds to $H_{\perp}=0$.}
\end{figure}

The short range attractive interaction between a vortex at position $x_i$ and the pinning potential
is approximated by $V[x_i(\tau)-x_o(\tau)]=-V_0 \delta[x_i-x_0(\tau)]$, where $x_0(\tau)$ is the pinning defect trajectory. The vortex-vortex interaction is assumed to be $U[x_i-x_j]=U_0\delta(x_i-x_j)$, with $U_0\rightarrow\infty$.
Upon assuming $N$ flux lines of length $2L_{\tau}$ confined to a length $L_x$ with periodic boundary conditions at $0$ and $L_x$ (see Fig.~\ref{fig:ManyLines}), we denote the positions of the vortices in the initial time-like slice $\tau_i$ as $\{x\}=(x_1,x_2,\dots,x_N)$ and in the final slice at $\tau_f$ as $\{x'\}=(x_1',x_2',\dots,x_N')$. 
The (classical) partition function for $N$ flux lines then reads:
\begin{equation}\label{PartitionSh2}
\begin{split}
\mathcal{Z}&\equiv\mathcal{Z}[\{x'\},\tau_f; \{x\},\tau_i;x_0(\tau)]\nonumber\\
&=\int \prod\limits_{i=1}^N\mathcal{D} x_i(\tau) \exp(-\frac{1}{T}\sum_i\int\limits_{\tau_i}^{\tau_f}\mathrm{d}\tau \Big[ \frac{\gamma}{2}\left(\frac{dx_i(\tau)}{d\tau}\right)^2+V[x_i(\tau)-x_0(\tau)]+\sum_j U[x_i(\tau)-x_j(\tau)]\Big])
\end{split}
\end{equation}
where $\gamma$ is the coarse-grained line tension. We continue to set Boltzmann's constant to $k_B\equiv 1$.

The transformation $x_i(\tau)\rightarrow y_i(\tau)=x_i(\tau)-x_0(\tau)$, $i=1,\dots,N$ for every vortex trajectory, enables us to switch to a frame of reference with a straight defect but varying externally imposed transverse magnetic field $H_{\perp}\propto h(\tau)$. We can now write the partition function of the many particle system as: 
\begin{equation}\label{PartitionSh}
 \mathcal{Z} = \mathcal{Z} [\{x'-x_o(\tau_f)\},\tau_f; \{x-x_0(\tau_i)\},\tau_i;h(\tau)] 
\end{equation}
where $h(\tau)=dx_0(\tau)/d\tau$, and the $h$-dependent analogue of Eq.~(\ref{PartitionSh2}) is given by Eq.~(\ref{PartitionSh3})

\begin{equation}\label{PartitionSh3}
\mathcal{Z}=\int \prod\limits_{i=1}^N\mathcal{D} x_i(\tau) \exp(-\frac{1}{T}\sum_i\int\limits_{\tau_i}^{\tau_f}\mathrm{d}\tau \Big[ \frac{\gamma}{2}\left(\frac{dx_i(\tau)}{d\tau}+h(\tau)\right)^2
+V[x_i(\tau)]+\sum_j U[x_i(\tau)-x_j(\tau)]\Big]).
\end{equation}

As discussed in Sec.~\ref{sec:Introduction}, we consider the $g=1$ Luttinger liquid limit and treat the vortex filaments
as hard core bosons, or, equivalently (by means of a Wigner-Jordan transformation) as non-interacting fermions.\footnote{See Ref.~\cite{Refael06}, and references therein.} However, we believe that the results will be qualitatively correct in cases where the average vortex separation $a$ is large compared to the London penetration depth $\lambda$ which determines the range of the inter-vortex interaction potential $U(x)$, where $x$ is the vortex separation.  
The problem of calculating the classical partition function (\ref{PartitionSh3}) can be reformulated and recast in quantum language by mapping the thermally fluctuating vortices onto fermions propagating in imaginary time. In this mapping $T\rightarrow \hbar$ and $\gamma\rightarrow m$~\cite{Blatter94}. 
The partition function then simply becomes the propagator
\begin{equation}\label{Propagator}
 \mathcal{Z} =\langle\{x'-x_0(\tau_f)\}|T_{\tau}\{e^{-\int_{\tau_i}^{\tau_f}d\tau\;\hat{\mathcal{H}}(\tau)/T}\}|\{x-x_0(\tau_i)\}\rangle
\end{equation}
 where $T_{\tau}$ is the time ordering operator and $\mathcal{H}(\tau)$ is the time-dependent Hamiltonian of the many particle system associated with Eq.~(\ref{PartitionSh3}) above 
\begin{equation}\label{Hamimany}
\hat{\mathcal{H}}(\tau)=-\int\mathrm{d}x\left(\frac{T^2}{2\gamma}\hat{\psi}_L\left(\frac{d}{dx}\right)^2\hat{\psi}_R+T h(\tau)\hat{\psi}_L\frac{d}{dx}\hat{\psi}_R\right)-V_0 \hat{\psi}_L(0)\hat{\psi}_R(0) .
\end{equation}
This Hamiltonian is non-Hermitian, the non-Hermitian term being generated by the galilean transformation in imaginary time.
We find it therefore advantageous to work with a non-Hermitian set of particle field operators
\begin{eqnarray}\label{PartFieldOp}
\hat{\psi}_L(x)&=&\sum_k \phi^k_L(x,h) \hat{c_k}^{\dagger}\nonumber\\
\hat{\psi}_R(x)&=&\sum_k \phi^k_R(x,h) \hat{c_k}. 
\end{eqnarray}
$\phi^k_L(x,h)$ and $\phi^k_R(x,h)$ are the left and right, respectively, eigenstates of the single particle Hamiltonian (i.e. neglecting interactions) for $h=$const indexed by a wavevector $k$~\cite{Hatano97,Hatano98}.
Note that, with this definition some usual relations for the particle field operators do not apply, as $\hat{\psi}_L^{\dagger}\ne \hat{\psi}_R$. However, anticommutation relations, such as:
\begin{equation}\label{Anticom}
\{ \hat{\psi}_R(x),\hat{\psi}_L(x')\}=\delta(x-x')
\end{equation}
do hold, and the vortex density is given by 
\begin{equation}\label{DensOp}
\hat{n}(x)=\hat{\psi}_L(x)\hat{\psi}_R(x).
\end{equation}

We will chose periodic boundary conditions in the $x$ dimension for $\phi^k_L(x,h)$ and $\phi^k_R(x,h)$, a choice which is appropriate when $N$ is odd.\footnote{This choice is imposed by the antisymmetry of fermions and can be easily understood if one considers the Slater N-particle determinant and moves one particle by one spatial period.}
In the absence of a pin, the density of vortices is a constant $\langle\hat{n}(x)\rangle=n_0=N/L_x$. When a (possibly $\tau$-dependent) pinning potential is present, 
the probability density reads:
\begin{equation}\label{Density}
\langle n(x) \rangle_{\tau} \equiv \frac{{}_L\langle\Psi(\tau)|\hat{n}(x)|\Psi(\tau)\rangle_R}
{{}_L\langle\Psi(\tau)|\Psi(\tau)\rangle_R}
\end{equation}
where $|\Psi(\tau)\rangle_R$ is the result of the evolution operator acting on the initial condition:
\begin{equation}\label{PsiR}
|\Psi(\tau)\rangle_R=T_{\tau}\{e^{-\int_{-L_{\tau}}^{\tau} \hat{\mathcal{H}}(\tau')\mathrm{d}\tau'}\}|\Psi^i\rangle_R,
\end{equation}
and similarly for ${}_L\langle\Psi(\tau)|$:
\begin{equation}\label{PsiL}
{}_L\langle\Psi(\tau)|={}_L\langle\Psi^f|T_{\tau}\{e^{-\int^{L_{\tau}}_{\tau} \hat{\mathcal{H}}(\tau')\mathrm{d}\tau'}\}.
\end{equation}
To simplify notation we now also set $T=1$.
Here, $|\Psi^i\rangle_R$ and ${}_L\langle\Psi^f|$ are non-hermitian generalizations of the usual filled Fermi sea ground state in one dimension (see below).
As in Fig.~\ref{fig:ManyLines}, $L_{\tau}$ is the half length of the slab in the time-like direction.

\begin{figure}\label{fig:fermisea}
\includegraphics[scale=1.]{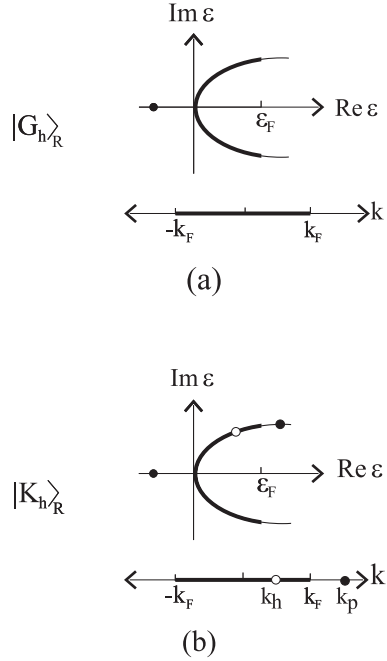}
\caption{Energy spectra and momenta for many body states for $N\rightarrow\infty$ when $h<h_c=V_0/T$. The isolated dot on the $\mathrm{Re}\epsilon$ axis represents the bound single particle state and does not appear on the wavevector line, but can participate in a particle-hole pair. (a) Ground state. (b) Single particle-hole excitation state.}
\end{figure}

To make the calculation of $\langle n(x) \rangle_{\tau} $ analytically tractable, we now consider the simple example of a defect trajectory with a kink at $\tau=0$: $h(\tau)=h \Theta(-\tau)+h' \Theta(\tau)$, with two slopes $h$ and $h'$, 
\begin{equation}\label{xotraje}
x_0(\tau)= h \tau \Theta(-\tau) + h' \tau \Theta(\tau).
\end{equation}
Moreover, we assume ground state boundary conditions, namely: $|\Psi^i\rangle=|G_h\rangle_R$ and $\langle\Psi^f|={}_L\langle G_{h'}|$, where
$|G_h\rangle_R$ is the right $N$ particle ground state of $\mathcal{H}(\tau)$ for $h(\tau)=h$ and
${}_L\langle G_{h'}|$ is the left $N$ particle ground state of $\hat{\mathcal{H}}(\tau)$ for $h(\tau)=h'$.
The above assumption is not unreasonable if we take $L_{\tau}\pm\tau\gg 1$ so the system has "time" to relax to the ground state both before and after it approaches the kink. Here $|G_h\rangle_R$ is the filled fermi sea ground state constructed from the eigenvalues of Ref.~\cite{Hatano97}. It includes the bound state (see Ref.~\cite{Hatano97}), represented as a dot on the $\mathrm{Re}\epsilon$ axis of the energy spectrum diagram in Fig.~\ref{fig:fermisea}(a) and the extended state that occupy the paraboloid part of the spectrum.

To compute the probability density distribution of the flux lines for $\tau>0$, we rewrite the time evolution operator as 
\begin{equation}
T_{\tau'}\{e^{-\int_{-L_{\tau}}^{\tau}d\tau'\;\hat{\mathcal{H}}(\tau')}\}=e^{-\tau\;\hat{\mathcal{H}}(\tau'>0)}e^{-L_{\tau}\;\hat{\mathcal{H}}(\tau'<0)}
\end{equation}
and insert a complete set of (normalized) many body eigenstates $\hat{I}=\sum\limits_{K_{h'}}|K_{h'}\rangle_R\;{}_L\langle K_{h'}|$. 
After some straightforward algebra we get:
\begin{equation}\label{DensExp}
\langle n(x) \rangle_{\tau}=\frac{1}{\mathcal{R}}\sum\limits_{K_{h'}} {}_L\langle G_{h'}|\hat{n}(x)|K_{h'}\rangle_R
{}_L\langle K_{h'}|G_{h}\rangle_R \; e^{-\tau[E_K(h')-E_G(h')]}
\end{equation}
where $\mathcal{R}$ is a normalization constant equal to
\begin{equation}\label{Norm}
\mathcal{R}={}_L\langle G_{h'}|G_h\rangle_R, 
\end{equation}
and the sum is over all possible $N$ particle eigenstates of $\mathcal{H}(\tau>0)$. Here $E_K(h')$ is the energy of the many body state $|K_{h'}\rangle_R$, and similarly $E_G(h)$ is the energy of ${}_L\langle G_{h'}|$. Note that in general $E_K(h)$ has both a real and imaginary part, since the Hamiltonian it corresponds to is non-Hermitian and that although $(|K_h\rangle_R)^{\dagger}\ne{}_L\langle K_h|$, these left and right eigenstates have the same eigenenergy. 

Since $\hat{n}(x)$ is a single particle operator, the matrix element ${}_L\langle G_{h'}|\hat{n}(x)|K_{h'}\rangle_R$ is zero for excited states with two or more particle-hole excitations. 
The summation is over all $K_{h'}$ states, where
\begin{equation}
|K_{h'}\rangle_R=|-k_F,-k_F+\frac{2\pi}{L_x}, ...,k_h-\frac{2\pi}{L_x},k_p, k_h+\frac{2\pi}{L_x},...,k_F\rangle.
\end{equation}
$k_F$ is the Fermi momentum, which equals $k_F=\frac{\pi}{L_x}(N-2)\simeq\pi n_0$ for $h<h_c$. Although not explicitly shown in this notation, the bound single particle eigenstate is also included in the particle-hole excitation summation.  
These states correspond to single particle-hole excitations of the filled fermi sea $|G_{h'}\rangle_R$ as shown in Fig.~\ref{fig:fermisea}.

Eq.~(\ref{DensExp}) effectively becomes a summation over all hole momenta $|k_h|\le k_F$ and all particle momenta $|k_p|> k_F$.

The non-Hermitian particle field operators defined earlier lead to the matrix element:
\begin{displaymath}\label{DensEl}
{}_L\langle G_{h'}|\hat{n}(x)|K_{h'}\rangle_R=\left\{ \begin{array}{ll} \phi^p_L(x,h') \phi^h_R(x,h')&\textrm{for $K_{h'}\ne G_{h'}$}\nonumber\\
\sum_n\phi^n_L(x,h') \phi^n_R(x,h')&\textrm{for $K_{h'}=G_{h'}$}
\end{array}\right.
\end{displaymath}
where $ \phi^p_R(x,h')$,  $\phi^h_L(x,h')$ are the single particle wavefunctions of the particle-hole pair and the $n$ summation is over all occupied single particle eigenstates in $G_h$.
It is easy to see that ${}_L\langle K_{h'}|G_{h}\rangle_R=\mathrm{det}(\tilde{C})$, where the matrix $\tilde{C}$ has elements $C_{i\;k}\equiv \int\limits_{-L_x/2}^{L_x/2} \mathrm{d}x \phi^L_i(x,h') \phi^R_k(x,h)$. 
\begin{figure}

\resizebox{0.5\textwidth}{!}{
\includegraphics{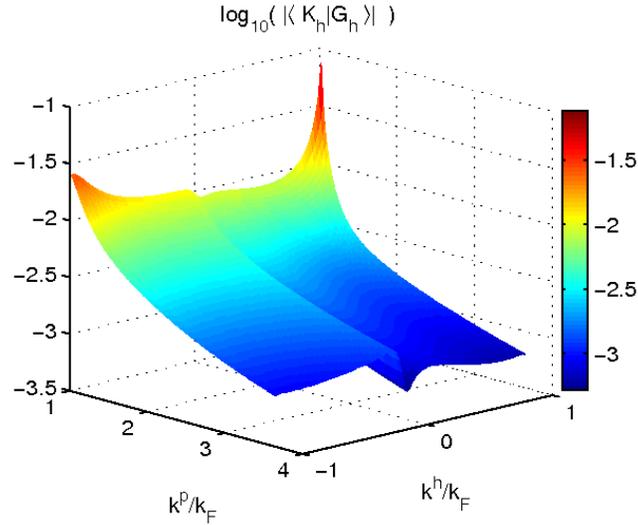}
}
\caption{Logarithm (base ten) of the absolute value of the matrix element $ {}_L\langle K_{h'}|G_{h}\rangle_R$ where $K_{h'}$ differs from the ground state $G_{h'}$ by a single particle-hole excitation with hole momentum $|k^h|<k_F$ and particle momentum $|k_p|>k_F$. $ {}_L\langle K_{h'}|G_{h}\rangle_R$ is maximum for $|k_p|\rightarrow k_F^+$ and at $k_p\simeq3k_F$ it has decreased by more than 2 orders of magnitude.}
\label{fig:MatrixElement}
\end{figure}

$ {}_L\langle K_{h'}|G_{h}\rangle_R$ becomes negligible for large $k_p$ momenta, which allows us to introduce a cutoff $k_c$ at the summation in Eq.~(\ref{DensExp}). An example is shown in Fig.~\ref{fig:MatrixElement}, where we plot the logarithm of $ {}_L\langle K_{h'}|G_{h}\rangle_R$ for a non-Hermitian field that changes from $h=0.05$ to $h'=0.5$ and an average vortex density $n_0=0.25$. 
The value $h=0.05$ was chosen instead of $h=0$ for the $\tau<0$ tilt, due to the implicit assumptions used in the evaluation of the single particle eigenstates that appear in Eq.~(\ref{PartFieldOp}): To allow for analytical results in closed form, $\phi_L^k$ and $\phi_R^k$ were derived under the assumption $e^{-L_x h \gamma/T}\ll 1$, which for finite systems is violated as $h\rightarrow 0$. The small initial tilt does not qualitatively change our observations. 

Having calculated $\langle n(x) \rangle_{\tau}$, we can undo our transformation, $x\rightarrow x-x_0(\tau)$ to get the actual probability density in the original frame of reference: 
\begin{equation}\label{DensPos}
\langle n(x) \rangle_{\tau}=\frac{1}{\mathcal{R}}\sum_{\substack{k_F<|k_p|<k_c\\|k_h|<k_F}} \phi^h_L(x-x_0(\tau),h') \phi^p_R(x-x_0(\tau),h')
\langle {}_L K_{h'}|G_{h}\rangle_R e^{-\tau(\epsilon_p(h')-\epsilon_h(h'))}+n_{h'}(x-x_0(\tau))
\end{equation}
$\epsilon_p(h')-\epsilon_h(h')$ is the particle-hole energy and $n_{h'}(x)$ is the vortex density $\langle n(x,\tau) \rangle$ for constant $h(\tau)=h'$.

We can easily adapt the above treatment to get the probability density for $\tau<0$:
\begin{equation}\label{DensNeg}
\langle n(x) \rangle_{\tau}=\frac{1}{\mathcal{R}}\sum_{\substack{k_F<|k_p|<k_c\\|k_h|<k_F}} \phi^p_L(x-x_0(\tau),h) \phi^h_R(x-x_0(\tau),h)\; {}_L\langle G_{h'}|K_{h}\rangle_R e^{\tau(\epsilon_p(h)-\epsilon_h(h))}+n_h(x-x_0(\tau))
\end{equation}

Another quantity which we can give us some insight to the properties of the transition between the $h$ and $h'$ sections of the tilted defect is the expectation value of the current operator ~\cite{Affleck04,Hatano97}:
\begin{equation}\label{Current}
\hat{J}=-i\frac{d\mathcal{H}(\tau)}{dh}=\frac{i T}{2}\int\left[\hat{\psi}_L(\frac{d}{dx}\hat{\psi}_R)-(\frac{d}{dx}\hat{\psi}_L)\hat{\psi}_R\right]dx
\end{equation}
Note that although the Hamiltonian depends on $h$, the current operator itself is explicitly $h$-independent.
With our definition, $\hat{J}$ is not a current density operator, but an integrated, position independent quantity, a measure of the total transverse magnetization due to the tilted flux lines.
 
In the absence of a pinning defect, the current $\langle\hat{J}\rangle$ is zero since the flux lines are parallel to the inducing magnetic field. When a defect is present, the flux lines bend in the vicinity of the tilted defect, thus creating a non-zero $h$-dependent current. As long as the magnetic field remains parallel to the $\tau$-axis, the total current $\langle J \rangle$ is independent of the system size, since vortices far from the defect are parallel to the magnetic field and do not contribute to the current.

\section{Oscillations in the time-like dimension}
\label{sec:Oscill}

In this Section we present some results obtained using the formalism discussed in Sec.~\ref{sec:ProbDistr}, and examine the flux line configuration in the vicinity of the defect kink.

\begin{figure}

\resizebox{0.5\textwidth}{!}{
\includegraphics{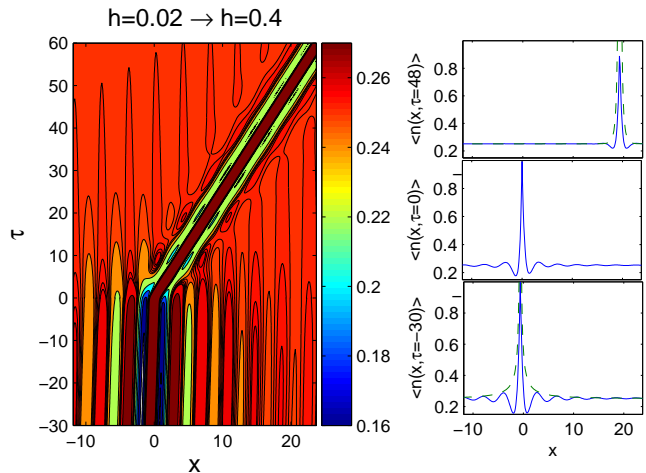}
}
\caption{Density of vortices when the tilt of the linear defect changes from $h=$0.02 to $h'=$0.4. For better visualization, the range of the $n(x,\tau)$ values was restricted to $[0.16, 0.27]$. On the right: 
snapshots for times $\tau=$-30, $\tau=$0 and $\tau=$48. Note the Friedel oscillations. The dashed curves represent the $\exp(-|x|/\xi)/|x|$ envelope of Eq.~(\ref{FriedelOsc}). Close to $\tau=0$, an assymetry develops in the $n(x,\tau)$ profile, and therefore such an envelope cannot be defined. Lengths are measured in units of $[x]=T^2/(\gamma V_0)$ and imaginary time in units of $[\tau]=T^3/(\gamma V_0^2)$.}
\label{fig:Density04}
\end{figure}

\begin{figure}

\resizebox{0.5\textwidth}{!}{
\includegraphics{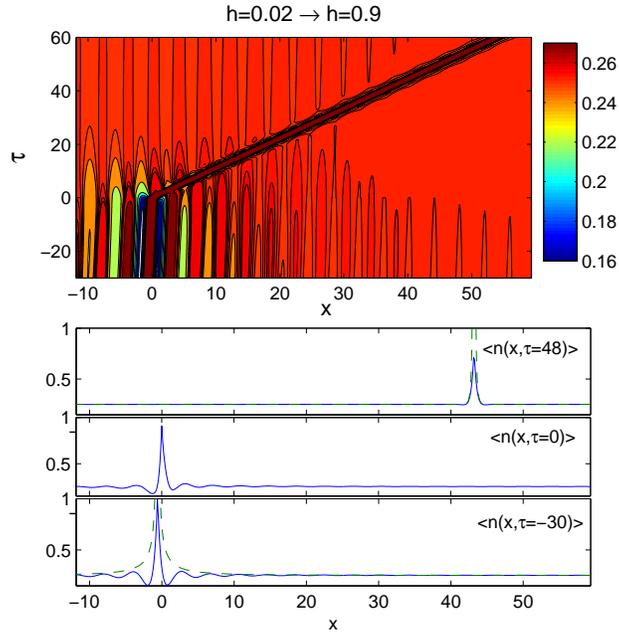}
}
\caption{Above: Density of vortices when the tilt of the linear defect changes from $h=$0.02 to $h'=$0.9. For better visualization, the range of the $n(x,\tau)$ values was restricted to $[0.16, 0.27]$. Below: 
snapshots for times $\tau=$-30, $\tau=$0 and $\tau=$48. The dashed curves on the top and bottom graphs represent the $\exp(-|x|/\xi)/|x|$ envelope of Eq.~(\ref{FriedelOsc}). $\xi$ is inversely proportional to $h$, so for high tilts no Friedel oscillations can be seen. Lengths are measured in units of $[x]=T^2/(\gamma V_0)$ and imaginary time in units of $[\tau]=T^3/(\gamma V_0^2)$.}
\label{fig:Density09}
\end{figure}

 Figures \ref{fig:Density04} and \ref{fig:Density09} show the vortex density distribution for a system of $N=91$ vortices penetrating a planar slab of length $L_x=360$ (measured in units of $T^2/\gamma V_0$), and interacting with a linear defect with tilt $h=0.02$ for $\tau<0$ and $h=0.4$ and $0.9$ respectively for $\tau>0$.

\begin{figure}

\resizebox{0.4\textwidth}{!}{
\includegraphics{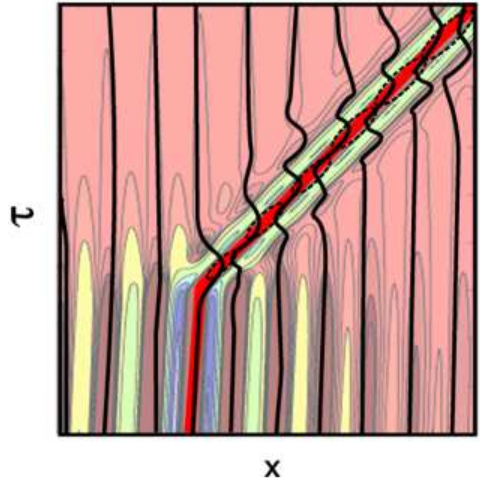}
}
\caption{ A probable configuration of flux lines on the planar superconductor. In this sketch, flux lines have been drawn to roughly follow the probability density maxima. The dotted ellipses demonstrate the likely position on the defect for the vortex exchange.}
\label{fig:VorticesWithBack}
\end{figure}

As $\tau\rightarrow\infty$, the probability distribution is peaked on the defect and exhibits oscillatory behavior with amplitude that decays as a power law modulated by an exponential, as described by Eq.~(\ref{FriedelOsc}) with $\alpha=1$ ("Friedel" oscillation behavior~\cite{Hofstetter04,Affleck04}). For $h \simeq 0$ the vortices form a periodic array, centered on the defect. The situation changes drastically close to the 
defect kink, especially for $\tau>0$. In this case there is a competition between the externally imposed magnetic
field $H$, which alone would result in an array of vortices parallel to the field, and the tilted defect.
The vortex which was localized on the defect for $\tau<0$ follows the defect until forced out by another vortex, which takes its place. Since the average vortex-vortex distance is $a=1/n_0$, the localized vortex stays on the defect over a time-like distance $\Delta\tau\sim 1/(n_0 h)$.
The exchange of vortices that are localized on the defect (one vortex enters while the next escapes the defect) takes place at the neighborhood of the probability density local minima $n(x_0(\tau_t),\tau_t)$. However, as $\tau\rightarrow\infty$, $n(x_0(\tau_t),\tau_t)$ becomes constant so the coordinate $\tau_t$ of the exchange is well defined only close to the kink. We expect that the $\tau$-independent probability distribution $n_{h'}(x,\tau\rightarrow\infty)$ is approached exponentially fast. A typical vortex configuration is presented in 
Fig.~\ref{fig:VorticesWithBack}.

\begin{figure}

\resizebox{0.4\textwidth}{!}{
\includegraphics{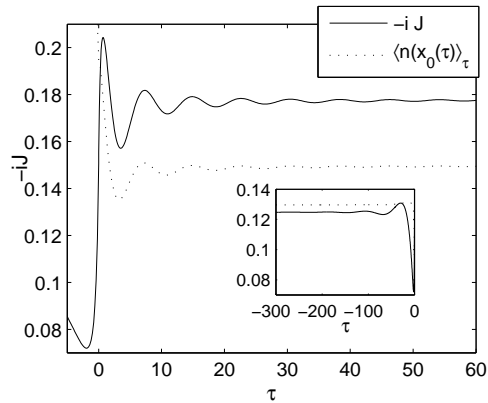}
}
\caption{Total imaginary current (see Eq.~(\ref{Current})) for $\tau>0$ transition $h=0.05\rightarrow h'=0.5$. The dotted line is a plot of $\langle n(x_0(\tau))\rangle_{\tau}$, the probability density on the defect in arbitrary units. Inset: Imaginary current for $\tau<0$.}
\label{fig:Current} 
\end{figure}

The kink breaks time-translational invariance and leads an  oscillatory behavior in the current, as well as the density, with period $\Delta\tau'=1/(n_0h')$. 
The imaginary component of the "current" is maximized when 
the vortex follows the defect, and minimized at $\tau_t$, the exchange point. The oscillations in the current 
die out far from the kink, as the ``jumps'' now occur with equal probability at all times $\tau$ and a traffic jam profile similar to the one described in Refs.~\cite{Hofstetter04,Affleck04} is formed. As can be seen in Fig.~\ref{fig:Current}, the maxima in the current coincide with the maxima of $n(x_0(\tau),\tau)$. An oscillatory behavior of the current also appears for $\tau<0$ as shown in the inset of Fig.~\ref{fig:Current}, this time with period $\Delta\tau=\frac{1}{n_0 h}$. $\Delta\tau$ diverges as $h\rightarrow 0$ and no oscillations in imaginary time are present. 

The length of the superconducting slab necessary to observe the phenomena discussed in this work would primarily depend on the imaginary time vortex-vortex collision period $\Delta\tau'\simeq a/h'$. If we take $U(x)$ to be proportional to the modified Bessel function of zeroth order (see for example \cite{Chorng98} and references therein), then for $a\simeq 3.5\lambda$ we have $U(a)/U(\lambda)<0.05$, so we would expect that the behaviour of the vortex system at that density could reasonably be approximated with a non-interacting model. For $\lambda\sim 40$nm, then for a system of 10 vortices we would need a superconducting wafer of width $\sim 1.5\mu$m and length $\sim 1.5\mu$m$/h'$, well within the experimentally accessible region.

\begin{figure}

\resizebox{0.5\textwidth}{!}{
\includegraphics{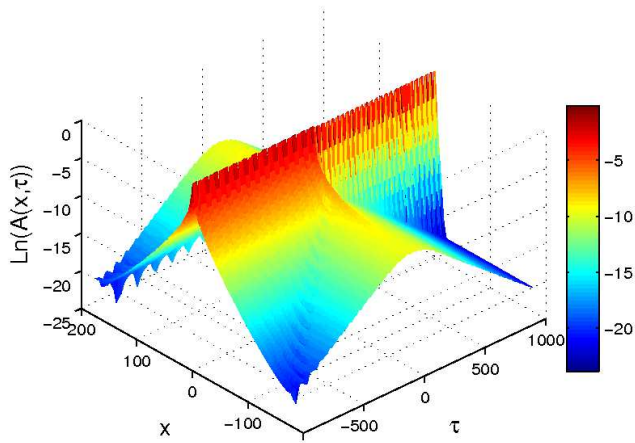}
}
\caption{ Logarithm of the amplitude describing the oscillatory decay of $n(x,\tau)$ away from a bent defect for a transition $h=0.05\rightarrow h'=0.5$, with $N=91$ vortices and $L_x=360$. The length and time units are $[x]=T^2/(\gamma V_0)$ and $[\tau]=T^3/(\gamma V_0^2)$ respectively. The frame of reference has been shifted so that the defect (and the amplitude maximum) remains at $x=0$ for both positive and negative $\tau$.}
\label{fig:Amplitude}
\end{figure}

The behaviors described before are summarized in Fig.~\ref{fig:Amplitude}, where we plot the logarithm of a local amplitude $A(x,\tau)$ of the oscillations for a system of $N=91$ vortices (again spread over a length $L_x=360$), in the presence of a defect with $h=0.05$ for $\tau<0$ and $h=0.5$ for $\tau>0$. 
The amplitude is a coarse-grained oscillation envelope obtained by subtracting the density of a local minimum from the density at the previous local maximum: $A((x_{\mathrm{min}}+x_{\mathrm{max}})/2,\tau)=n(x_{\mathrm{max}},\tau)-n(x_{\mathrm{min}},\tau)$.
For easier visualization, in this picture we have shifted the frame of reference to one which moves with the defect, so that the center of the defect is at $x=0$. 
As always, the defect changes slope at $\tau=0$.

The picture consists of an almost uniform in $x$ oscillatory background with a maximum at $\tau=0$ which decreases exponentially fast in $\tau$ (appearing as a straight line in a logarithmic plot) far from the kink.\footnote{As $L_x\rightarrow\infty$, the oscillatory background decays slowly to zero far from the defect.}
The oscillatory background is solely due to the presence of the kink in the defect trajectory. 
This abrupt change results in enhanced positional order of the vortices far from the defect. 
The "tent"-like structure emerging from the background as $\tau\rightarrow\pm\infty$ is due to the oscillatory behavior of the time independent part of the density $n_h(x)$ for $\tau<0$ and $n_{h'}$ for $\tau<0$. 
As the jump from $h$ to $h'$ at $\tau=0$ is approached, the width of the "tent"-like structure, which is a measure of the coherence length $\xi\sim 1/h$, decreases and the "tent" becomes narrower. 
Of course we should bear in mind that the quantity $A(x,\tau)$ is not well defined close to $x=0$, where $n(x,\tau)$ changes much faster than $\lambda\sim L_x/N$, the wavelength of he Friedel oscillations.
With this caveat, the oscillatory behavior at $x=0$ with period $\Delta\tau$ for $\tau<0$ and $\Delta\tau'$ for $\tau>0$ can be understood as a signature of the vortex interactions mediated by the kink in the defect.  

\begin{figure}

\resizebox{0.4\textwidth}{!}{
\includegraphics{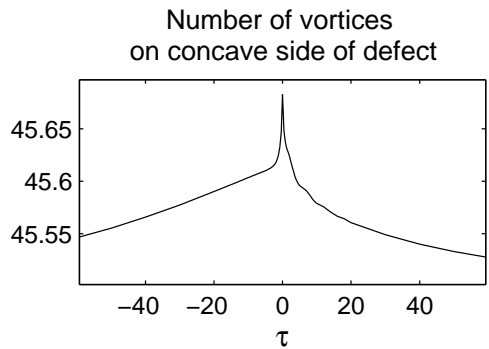}
}
\caption{ Total number of vortices $N_{\mathrm{c}}=\int^{x_0(\tau)+L_x/2}_{x_0(\tau)} \langle n(x) \rangle_{\tau} \mathrm{d}x$ on the concave side of the defect for a transition $h=0.02\rightarrow h'=0.7$.}
\label{fig:TotalVor}
\end{figure}

Far from the kink, at each time slice the vortex density distribution is symmetric about the defect position $x_0(\tau)$. Although not evident at the resolution of Fig.~\ref{fig:Amplitude}, the reflection symmetry of $n(x,\tau)$  across the defect is broken near the kink is broken. The density of vortices is higher on the concave side of the defect, although only by a small amount, as seen in Fig.~\ref{fig:TotalVor}. 
This effect was also present in the single vortex- meandering defect system, and disappears when $h'\rightarrow h$~\cite{Katifori06}.

\begin{figure}

\resizebox{0.4\textwidth}{!}{
\includegraphics{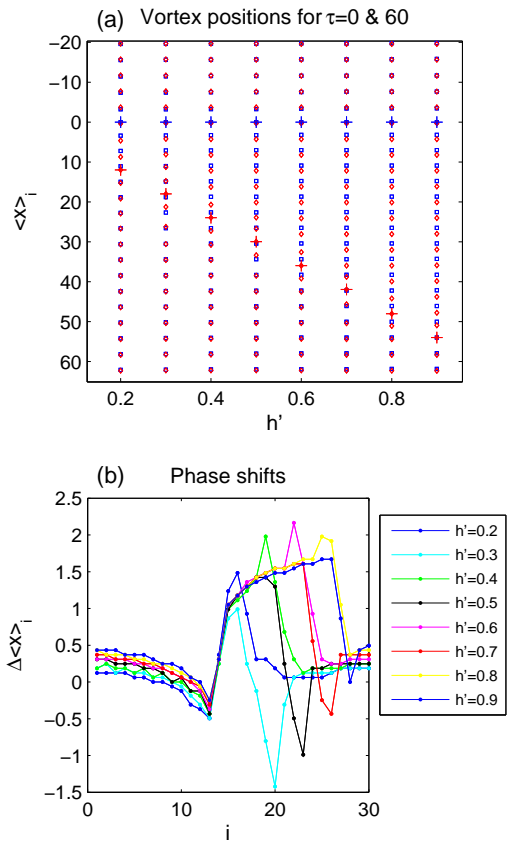}
}
\caption{(a)Average position flux lines at $\tau=0$ (squares) and at $\tau=60$ (diamonds). 
Each column corresponds to a different value of $h'$. In all cases, $h=0.02$. The blue cross corresponds to the position of the defect at $\tau=0$ and the red cross
at $\tau=60$. (b)Shift in average position of the vortices  $\langle x\rangle_i$ versus vortex number $i$. The vortices that cross the defect exhibit an approximately
constant shift in position, idependent of $h'$. }
\label{fig:phaseshifts}
\end{figure}

The average vortex positions $\langle x\rangle_i$ (corresponding to the local maxima of the vortex density) far from the defect kink follow the well known Friedel oscillation pattern ~\cite{Hofstetter04,Affleck04}, 
forming a periodic array with lattice spacing $1/n_0$. The phase of the oscillations far from the defect, depends on the tilt $h$ and the strength of the attractive
potential. However, as the defect kink is approached the simple periodic pattern in the vortex positions breaks down, and 
a phase shift is introduced. To explore how the defect kink affects $\langle x\rangle_i$ 
in Fig.~\ref{fig:phaseshifts}(a)  we plot the local maxima (average position of each flux line) of the probability distribution at imaginary time $\tau=0$ (squares) and at $\tau=60$ (diamonds). 
Each column corresponds to a different value of $h'$. In all cases, $h=0.02$. The blue cross corresponds to the position of the defect at $\tau=0$ and the red cross
at $\tau=60$.

At each column, the data points beetween two crosses correspond to the vortices whose trajectories crossed the defect between $\tau=0$ and $\tau=60$. By simple inspection
it can be determined that there is a minimal change in the average vortex position between $\tau=0$ and $\tau=60$ for vortices that did not cross the defect.
However, vortices that cross the defect get trapped and move with it for some time until forced out by the next vortex that impinges upon the defect. 

A more detailed description is given by Fig.~\ref{fig:phaseshifts}(b), where we plot the shift in average position of the vortices  $\langle x\rangle_i$ versus
vortex number $i$ for different tilts. We arbitrarily labeled the vortex at distance $\langle x \rangle\simeq 50$ as vortex $1$.
A positive $\Delta \langle x\rangle_i$ corresponds to a vortex position shifting to the right at $\tau=60$ with respect to its original position at $\tau=0$.
For all tilts $h'$, the vortex which is immediately to the left of the defect is shifted to the left. As discussed previously, vortices enter, become trapped and then exit the defect at 
periodic intervals. For certain tilts $h'$, the imaginary time slice $\tau=60$ happens to be when a vortex enters the defect, by abruptly turning to the left. This is the origin of the 
maxima that can be observed on Fig.~\ref{fig:phaseshifts}(b).

Ignoring those $\tau$ dependent features, all displacements $\Delta \langle x\rangle_i$ of the vortices that cross the defect collapse on the same value, a trapping displacement $l_{\mathrm{tr}}$ approximately equal to $l_{\mathrm{tr}}\simeq 0.38 a$, where $a$ is the mean vortex spacing. The trapping displacement does not depend on the tilt $h'$, as similarly defined quantities for single vortex systems, i.e. the trapping length in \cite{Hatano97} and the vortex shift $m(h)$ in \cite{Katifori06}, do. This contradicts the intuitive expectation, derived from the single pinned flux lines that $l_{\mathrm{tr}}$ should decrease as $h'$ increases.

\section{Potential experiments and defect lines that terminate}
\label{sec:Experiments}

\begin{figure}

\resizebox{0.4\textwidth}{!}{
\includegraphics{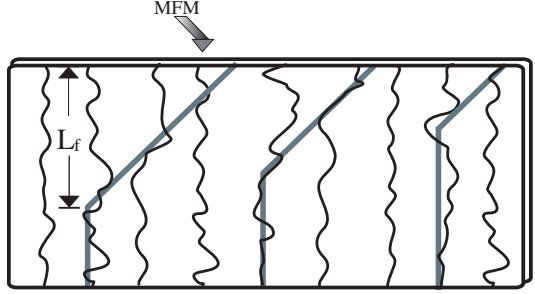}
}
\caption{ Sketch of possible experimental setup for the kinked defect system. The defects are represented with dark gray lines. 
The MFM tip probes the neighborhood of the defect exit position.}
\label{fig:expe}
\end{figure}

The above results and discussion about the probability density are applicable only far from the edges of the slab, a region difficult to probe experimentally. The use of ground state initial and final boundary conditions to derive the probability density in the core of the slab requires that the coordinate $\tau$ of interest be far from the boundaries. 
The experimentally relevant boundary condition (final or initial state) is an extension of the single particle boundary conditions $\int \mathrm{d}x \langle x|$ for many particles, namely $\langle \Psi^f|=\int\mathrm{d}x_1\int^{x_1}\mathrm{d}x_2\dots\int^{x_{N-1}}\mathrm{d}x_N\langle x_1,x_2,\dots,x_N|$ (and similarly for the initial state).
If we examine the probability density distribution at the upper boundary (which describes the exit positions of the vortices ), we can assume ground state boundary conditions at the other boundary provided $L_{\tau}\gg 1$. The energy spectrum of the single particle states has an imaginary component.  Due to this component, irrespective of the exact form of the exiting boundary conditions, we expect to observe oscillatory behavior with respect to the distance between the kink and the exit surface $L_f$.
The probability density at the upper boundary reads:
\begin{equation}\label{DensityFreeBc}
\langle n(x)\rangle_{L_f}=\frac{\sum_{K_{h'}} \langle\Psi^f|\hat{n}(x)|K_{h'}\rangle e^{-\Delta E(K_{h'})L_f} \langle K_{h'}|G_h\rangle} 
                             {\sum_{K_{h'}} \langle\Psi^f|K_{h'}\rangle e^{-\Delta E(K_{h'})L_f} \langle K_{h'}|G_h\rangle}
\end{equation}
where $\Delta E(K_{h'})=E(K_{h'})-E(G_{h'})$.
If, however, $L_f\gg 1$, then we can approximate the sum in the denominator of Eq.~(\ref{DensityFreeBc}) by the term $K_{h'}=G_{h'}$ and find for the density of exit points:
\begin{equation}\label{ExitOsci}
\langle n(x)\rangle_{L_f}\simeq n_{\Psi^f}(x)+\frac{1} {\langle\Psi^f|G_{h'}\rangle  \langle G_{h'}|G_h\rangle}\; \sum_{K_{h'}\ne G_{h'}} (\langle\Psi^f|\hat{n}(x)|K_{h'}\rangle- n_{\Psi^f}(x) \langle\Psi^f|K_{h'}\rangle)\; e^{-\Delta E(K_{h'})L_f} \langle K_{h'}|G_h\rangle
\end{equation}
Here, $n_{\Psi^f}(x)$ is the probability density distribution at the boundary for $L_f\rightarrow\infty$:
\begin{equation}\label{DensityFreeInf}
n_{\Psi^f}(x) \equiv \frac {\langle\Psi^f|\hat{n}(x)|G_{h'}\rangle}{\langle\Psi^f|G_{h'}\rangle}
\end{equation}
For sufficiently large $L_f$, the sum in the second term of Eq.~(\ref{ExitOsci}) is dominated by the lowest energy eigenstates $K_{h'}$. These are the single particle excitations with energy
$\mathrm{Re}E(K_{h'})\simeq N/L_x^2$ and momentum $\Delta p\simeq 2 k_F$ (the equality holds for $L_x\rightarrow\infty$). 
Note that $e^{-\Delta E\; L_f/T}\simeq e^{-i2 h k_F L_f}\simeq e^{-i 2\pi n_0 h L_f} $. Thus, if we keep $\Delta x=x-x_0(\tau)$, the distance from the defect, constant we expect to observe a periodic modulation in the $\tau$ direction of the vortex probability distribution. 
This periodicity  should manifest itself in magnetic force microscope (MFM) experiments as it traverses the slab \cite{Wadas92,Volodin98}, since the force necessary to pull the vortex from its exit position would depend on the length $L_f$ 
of the tilted segment of the defect. For example many linear defects with varying tilted segment length can be etched on the same superconducting slab, as 
in Fig.~\ref{fig:expe}. If the defects are sufficiently far apart so that we can assume that each vortex interacts with only one defect, then we expect $n(x,\tau)$
of the single defect system to approximate well the density at the neighborhood of each defect. With the proper choice of the tilted segments of the linear defects 
imprinted on the slab, one should be able to observe a $\sin(2\pi n_0 h L_f)$ dependence of the probability density at a fixed distance from each defect end point.

\begin{figure}

\resizebox{0.5\textwidth}{!}{
\includegraphics{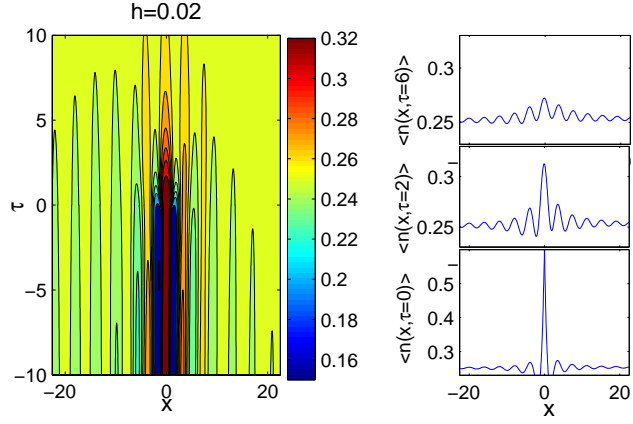}
}
\caption{ Density of vortices with a terminating linear defect with tilt $h=$0.02. The defect terminates at $\tau=0$.
On the right: snapshots for times $\tau=$0, $\tau=$2 and $\tau=$6. The slight assymetry of the density profile near $\tau=0$ is due to the nonzero tilt $h$ of the defect, and disappears for $h\rightarrow 0$. The density distribution delocalises exponentially fast after $\tau=0$. Lengths are measured in units of $[x]=T^2/(\gamma V_0)$ and imaginary time in 
units of $[\tau]=T^3/(\gamma V_0^2)$.}
\label{fig:xray02}
\end{figure}

\begin{figure}

\resizebox{0.5\textwidth}{!}{
\includegraphics{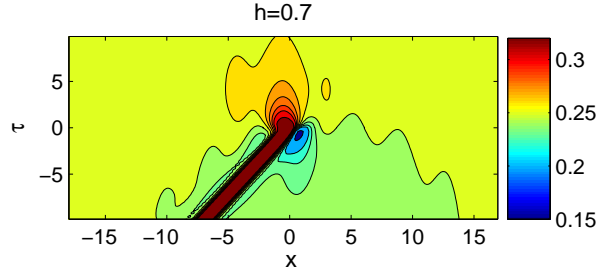}
}
\caption{ Density of vortices with a terminating linear defect with tilt $h=0.7$. No oscillatory behaviour in imaginary time is present.}
\label{fig:xray7}
\end{figure}

An interesting question that can be explored through a similar approach is the case of a defect terminating while still inside the slab. In this case the localised bound state
that existed for $h$ smaller than $h_c$, the critical tilt for the delocalised transition disappears for $\tau>0$ and the spectrum consists only of extended states. 
The localised vortex starts diffusing as it approaches the defect end and spatial phase information is quickly lost (Fig.~\ref{fig:xray02}). The "locking" of phase in the 
time-like dimension that was observed for kinked defects and the oscillations in $\tau$ of the vortex density that this resulted to are not present in the case of the 
terminating defect. This can be seen in Fig.~\ref{fig:xray7}, where the defect is at tilt $h=0.7$ with respect to the externally imposed field. Vortices still enter and exit 
the defect forming the traffic jam discussed in Refs.~\cite{Hofstetter04} and  ~\cite{Affleck04}, but the exit position of the last trapped vortex can vary so oscillations of the vortex density in the time like dimension are not 
observed.

\begin{acknowledgements}
We would like to thank M. Aziz, J. Blakely, V. Ignatesku and E. Williams for useful conversations. This work was supported by
the National Science Foundation through NSF Grant DMR-0231631 and the Harvard Materials Research Laboratory 
through NSF Grant NMR-0213805.
\end{acknowledgements}

\appendix
\section{Derivation of the non-Hermitian Hamiltonian}
The mapping of the classical flux line system to imaginary bosons (or fermions) is explained in many publications, see for example Refs.~\cite{Hatano97,Katifori06}. However for completeness, we briefly review it here. The partition function for one flux line interacting with a meandering defect is:
\begin{equation}\label{app_parttrans}
 \mathcal{Z}[ x_f,\tau_f;x_i,\tau_i;h(\tau)]=\int_{x(\tau_i)=x_i}^{x(\tau_f)=x_f}\mathcal{D} x(\tau) e^{-E[x(\tau)]/T}
\end{equation}
where
\begin{equation}
E[x(\tau)]=\int\limits_{\tau_i}^{\tau_f}d\tau \left[ \frac{\gamma}{2}\left(\frac{dx(\tau)}{d\tau}\right)^2-V_o\:\delta (x-x_0(\tau))\right].
\end{equation}
$x_i$ and $x_f$ are the positions of the vortex at $\tau_i$ and $\tau_f$ respectively.
A change of variables $y(\tau)=x(\tau)-x_0(\tau)$ will transform Eq.~(\ref{app_parttrans}) to a functional integral over the variable $y(\tau)$ with energy:
\begin{equation}
E[y(\tau)]=\int\limits_{\tau_i}^{\tau_f}d\tau \left[ \frac{\gamma}{2}\left(\frac{dy(\tau)}{d\tau}+h(\tau)\right)^2-V_o\:\delta (y)\right]
\end{equation}
where $h(\tau)=dx_0(\tau)/d\tau$.

The partition function obeys the equation:
\begin{equation}
-\frac{\partial}{\partial\tau} \mathcal{Z}[ y,\tau;y_i,\tau_i;h(\tau)]=\big(-\frac{T^2}{2\gamma}\frac{\partial^2}{\partial y^2}-h(\tau) T \frac{\partial}{\partial y}\;-V_0\delta(y)\big)\mathcal{Z}[ y,\tau;y_i,\tau_i;h(\tau)].
\end{equation}

Thus, $\mathcal{Z}[ x,\tau;y_i,\tau_i;h(\tau)]$ can be thought of as the quantum mechanical propagator of a particle propagating in imaginary time, with non-Hermitian Hamiltonian:
\begin{equation}\label{AppHami}
\mathcal{H}(\tau)=\frac{1}{2\gamma}\hat{p}^2-i h(\tau)  \hat{p}-V_0\delta(x)
\end{equation}
where $\hat{p}\equiv \frac{T}{i}\frac{\partial}{\partial x}$.
The generalization to many lines is straightforward.

The non-Hermitian single particle eigenstates $ \phi^k_R(x,h)$ used throughout this work are derived from solving Eq.~(\ref{AppHami}) for
$h(\tau)=\mathrm{cons}$\cite{Hatano97}. When $h<h_c=V_0/T$, for a system of length $L_x$, with $x\in [0,L_x]$ and periodic boundary conditions, the (unnormalized) extended states read:
\begin{equation}
 \phi^k_R(x,h)= \frac{e^{-ikx+\delta\kappa x} }{e^{-ikL_x+\delta\kappa L_x} -1}
-\frac{e^{ikx-2 \gamma h x/T-\delta\kappa x} }{e^{ik L_x-2 \gamma h L_x/T-\delta\kappa L_x}-1}
\end{equation}
where
\begin{equation}
\delta\kappa=\frac{1}{L_x} \mathrm{ln} \left[ -\frac{k+i\frac{\gamma h}{T} }{k+i\frac{\gamma}{T}(h-\frac{V_0}{T})} \right]
\end{equation}
provided that $e^{-2L_x\gamma h /T}\ll 1$.
The bound state similalry reads
\begin{equation}
 \phi^0_R(x,h)= \frac{e^{\frac{\gamma}{T}(\frac{V_0}{T}-h) x} }{e^{\frac{\gamma}{T}(\frac{V_0}{T}-h) L_x}-1}
-\frac{e^{-\frac{\gamma}{T}(\frac{V_0}{T}+h) x} }{e^{-\frac{\gamma}{T}(\frac{V_0}{T}+h) L_x}-1}
\end{equation}


\begin{thebibliography}{23}
\expandafter\ifx\csname natexlab\endcsname\relax\def\natexlab#1{#1}\fi
\expandafter\ifx\csname bibnamefont\endcsname\relax
  \def\bibnamefont#1{#1}\fi
\expandafter\ifx\csname bibfnamefont\endcsname\relax
  \def\bibfnamefont#1{#1}\fi
\expandafter\ifx\csname citenamefont\endcsname\relax
  \def\citenamefont#1{#1}\fi
\expandafter\ifx\csname url\endcsname\relax
  \def\url#1{\texttt{#1}}\fi
\expandafter\ifx\csname urlprefix\endcsname\relax\def\urlprefix{URL }\fi
\providecommand{\bibinfo}[2]{#2}
\providecommand{\eprint}[2][]{\url{#2}}

\bibitem[{\citenamefont{Blatter et~al.}(1994)\citenamefont{Blatter, Feigel'man,
  Geshkenbein, Larkin, and Vinokur}}]{Blatter94}
\bibinfo{author}{\bibfnamefont{G.}~\bibnamefont{Blatter}},
  \bibinfo{author}{\bibfnamefont{M.~V.} \bibnamefont{Feigel'man}},
  \bibinfo{author}{\bibfnamefont{V.~B.} \bibnamefont{Geshkenbein}},
  \bibinfo{author}{\bibfnamefont{A.~I.} \bibnamefont{Larkin}},
  \bibnamefont{and} \bibinfo{author}{\bibfnamefont{V.~M.}
  \bibnamefont{Vinokur}}, \bibinfo{journal}{Rev.\ Mod.\ Phys.}
  \textbf{\bibinfo{volume}{66}}, \bibinfo{pages}{1125} (\bibinfo{year}{1994}).

\bibitem[{\citenamefont{Fisher et~al.}(1991)\citenamefont{Fisher, Fisher, and
  Huse}}]{Fish91}
\bibinfo{author}{\bibfnamefont{D.~S.} \bibnamefont{Fisher}},
  \bibinfo{author}{\bibfnamefont{M.~P.~A.} \bibnamefont{Fisher}},
  \bibnamefont{and} \bibinfo{author}{\bibfnamefont{D.~A.} \bibnamefont{Huse}},
  \bibinfo{journal}{Phys.\ Rev. B} \textbf{\bibinfo{volume}{43}},
  \bibinfo{pages}{130} (\bibinfo{year}{1991}).

\bibitem[{\citenamefont{Hwa et~al.}(1993{\natexlab{a}})\citenamefont{Hwa,
  Doussal, Nelson, and Vinokur}}]{Hwa93}
\bibinfo{author}{\bibfnamefont{T.}~\bibnamefont{Hwa}},
  \bibinfo{author}{\bibfnamefont{P.~L.} \bibnamefont{Doussal}},
  \bibinfo{author}{\bibfnamefont{D.~R.} \bibnamefont{Nelson}},
  \bibnamefont{and} \bibinfo{author}{\bibfnamefont{V.}~\bibnamefont{Vinokur}},
  \bibinfo{journal}{Phys. \ Rev. \ Lett.} \textbf{\bibinfo{volume}{71}},
  \bibinfo{pages}{3545} (\bibinfo{year}{1993}{\natexlab{a}}).

\bibitem[{\citenamefont{Nelson and Vinokur}(1993)}]{Nelson93}
\bibinfo{author}{\bibfnamefont{D.~R.} \bibnamefont{Nelson}} \bibnamefont{and}
  \bibinfo{author}{\bibfnamefont{V.~M.} \bibnamefont{Vinokur}},
  \bibinfo{journal}{Phys. Rev. B} \textbf{\bibinfo{volume}{48}},
  \bibinfo{pages}{13060} (\bibinfo{year}{1993}).

\bibitem[{\citenamefont{Bolle et~al.}(1999)\citenamefont{Bolle, Aksyuk, Pardo,
  Gammel, Zeldov, Bucher, Boie, Bishop, and Nelson}}]{Bolle99}
\bibinfo{author}{\bibfnamefont{C.~A.} \bibnamefont{Bolle}},
  \bibinfo{author}{\bibfnamefont{V.}~\bibnamefont{Aksyuk}},
  \bibinfo{author}{\bibfnamefont{F.}~\bibnamefont{Pardo}},
  \bibinfo{author}{\bibfnamefont{P.~L.} \bibnamefont{Gammel}},
  \bibinfo{author}{\bibfnamefont{E.}~\bibnamefont{Zeldov}},
  \bibinfo{author}{\bibfnamefont{E.}~\bibnamefont{Bucher}},
  \bibinfo{author}{\bibfnamefont{R.}~\bibnamefont{Boie}},
  \bibinfo{author}{\bibfnamefont{D.~J.} \bibnamefont{Bishop}},
  \bibnamefont{and} \bibinfo{author}{\bibfnamefont{D.~R.}
  \bibnamefont{Nelson}}, \bibinfo{journal}{Nature}
  \textbf{\bibinfo{volume}{399}}, \bibinfo{pages}{43} (\bibinfo{year}{1999}).

\bibitem[{\citenamefont{Tonomura et~al.}(2001)\citenamefont{Tonomura, Kasai,
  Kamimura, Matsuda, Harada, Nakayama, Shimoyama, Kishio, Hanaguri, Kitazawa
  et~al.}}]{Tonomura01}
\bibinfo{author}{\bibfnamefont{A.}~\bibnamefont{Tonomura}},
  \bibinfo{author}{\bibfnamefont{H.}~\bibnamefont{Kasai}},
  \bibinfo{author}{\bibfnamefont{O.}~\bibnamefont{Kamimura}},
  \bibinfo{author}{\bibfnamefont{T.}~\bibnamefont{Matsuda}},
  \bibinfo{author}{\bibfnamefont{K.}~\bibnamefont{Harada}},
  \bibinfo{author}{\bibfnamefont{Y.}~\bibnamefont{Nakayama}},
  \bibinfo{author}{\bibfnamefont{J.}~\bibnamefont{Shimoyama}},
  \bibinfo{author}{\bibfnamefont{K.}~\bibnamefont{Kishio}},
  \bibinfo{author}{\bibfnamefont{T.}~\bibnamefont{Hanaguri}},
  \bibinfo{author}{\bibfnamefont{K.}~\bibnamefont{Kitazawa}},
  \bibnamefont{et~al.}, \bibinfo{journal}{Nature}
  \textbf{\bibinfo{volume}{412}}, \bibinfo{pages}{620} (\bibinfo{year}{2001}).

\bibitem[{\citenamefont{Polkovnikov et~al.}(2005)\citenamefont{Polkovnikov,
  Kafri, and Nelson}}]{Polkovnikov05}
\bibinfo{author}{\bibfnamefont{A.}~\bibnamefont{Polkovnikov}},
  \bibinfo{author}{\bibfnamefont{Y.}~\bibnamefont{Kafri}}, \bibnamefont{and}
  \bibinfo{author}{\bibfnamefont{D.~R.} \bibnamefont{Nelson}},
  \bibinfo{journal}{Phys.\ Rev. B} \textbf{\bibinfo{volume}{71}},
  \bibinfo{pages}{014511} (\bibinfo{year}{2005}).

\bibitem[{\citenamefont{Refael et~al.}(2006)\citenamefont{Refael, Hofstetter,
  and Nelson}}]{Refael06}
\bibinfo{author}{\bibfnamefont{G.}~\bibnamefont{Refael}},
  \bibinfo{author}{\bibfnamefont{W.}~\bibnamefont{Hofstetter}},
  \bibnamefont{and} \bibinfo{author}{\bibfnamefont{D.~R.}
  \bibnamefont{Nelson}}, \bibinfo{journal}{Phys. \ Rev. B}
  \textbf{\bibinfo{volume}{74}}, \bibinfo{pages}{174520}
  (\bibinfo{year}{2006}).

\bibitem[{\citenamefont{Hwa et~al.}(1993{\natexlab{b}})\citenamefont{Hwa,
  Nelson, and Vinokur}}]{Hwa93b}
\bibinfo{author}{\bibfnamefont{T.}~\bibnamefont{Hwa}},
  \bibinfo{author}{\bibfnamefont{D.~R.} \bibnamefont{Nelson}},
  \bibnamefont{and} \bibinfo{author}{\bibfnamefont{V.}~\bibnamefont{Vinokur}},
  \bibinfo{journal}{Phys. \ Rev. B} \textbf{\bibinfo{volume}{48}},
  \bibinfo{pages}{1167} (\bibinfo{year}{1993}{\natexlab{b}}).

\bibitem[{\citenamefont{Devereaux et~al.}(1994)\citenamefont{Devereaux,
  Scalettar, and Zimanyi}}]{Derev94}
\bibinfo{author}{\bibfnamefont{T.~P.} \bibnamefont{Devereaux}},
  \bibinfo{author}{\bibfnamefont{R.~T.} \bibnamefont{Scalettar}},
  \bibnamefont{and} \bibinfo{author}{\bibfnamefont{G.~T.}
  \bibnamefont{Zimanyi}}, \bibinfo{journal}{Phys. \ Rev. B}
  \textbf{\bibinfo{volume}{50}}, \bibinfo{pages}{13625} (\bibinfo{year}{1994}).

\bibitem[{\citenamefont{Hofstetter et~al.}(2004)\citenamefont{Hofstetter,
  Affleck, Nelson, and Schollwock}}]{Hofstetter04}
\bibinfo{author}{\bibfnamefont{W.}~\bibnamefont{Hofstetter}},
  \bibinfo{author}{\bibfnamefont{I.}~\bibnamefont{Affleck}},
  \bibinfo{author}{\bibfnamefont{D.~R.} \bibnamefont{Nelson}},
  \bibnamefont{and}
  \bibinfo{author}{\bibfnamefont{U.}~\bibnamefont{Schollwock}},
  \bibinfo{journal}{Europhys.\ Lett. B} \textbf{\bibinfo{volume}{66}},
  \bibinfo{pages}{178} (\bibinfo{year}{2004}).

\bibitem[{\citenamefont{Affleck et~al.}(2004)\citenamefont{Affleck, Hofstetter,
  Nelson, and Schollwock}}]{Affleck04}
\bibinfo{author}{\bibfnamefont{I.}~\bibnamefont{Affleck}},
  \bibinfo{author}{\bibfnamefont{W.}~\bibnamefont{Hofstetter}},
  \bibinfo{author}{\bibfnamefont{D.~R.} \bibnamefont{Nelson}},
  \bibnamefont{and}
  \bibinfo{author}{\bibfnamefont{U.}~\bibnamefont{Schollwock}},
  \bibinfo{journal}{J. Stat. Mech.} \textbf{\bibinfo{volume}{10}},
  \bibinfo{pages}{P10003} (\bibinfo{year}{2004}).

\bibitem[{\citenamefont{Radzihovsky}(2006)}]{Radzihovsky06}
\bibinfo{author}{\bibfnamefont{L.}~\bibnamefont{Radzihovsky}},
  \bibinfo{journal}{Phys.\ Rev. B} \textbf{\bibinfo{volume}{73}},
  \bibinfo{pages}{104504} (\bibinfo{year}{2006}).

\bibitem[{\citenamefont{Pokrovsky and Talapov}(1979)}]{Pokrovsky79}
\bibinfo{author}{\bibfnamefont{V.~L.} \bibnamefont{Pokrovsky}}
  \bibnamefont{and} \bibinfo{author}{\bibfnamefont{A.~L.}
  \bibnamefont{Talapov}}, \bibinfo{journal}{Phys.\ Rev. \ Lett.}
  \textbf{\bibinfo{volume}{42}}, \bibinfo{pages}{65} (\bibinfo{year}{1979}).

\bibitem[{\citenamefont{Pokrovsky and Talapov}(1984)}]{Pokrovsky84}
\bibinfo{author}{\bibfnamefont{V.~L.} \bibnamefont{Pokrovsky}}
  \bibnamefont{and} \bibinfo{author}{\bibfnamefont{A.~L.}
  \bibnamefont{Talapov}}, \bibinfo{journal}{Soviet Scientific Reviews
  Supplement Series-Physics} \textbf{\bibinfo{volume}{1}}
  (\bibinfo{year}{1984}).

\bibitem[{\citenamefont{Coppersmith et~al.}(1982)\citenamefont{Coppersmith,
  Fisher, Halperin, Lee, and Brinkman}}]{Coppersmith82}
\bibinfo{author}{\bibfnamefont{S.~N.} \bibnamefont{Coppersmith}},
  \bibinfo{author}{\bibfnamefont{D.~S.} \bibnamefont{Fisher}},
  \bibinfo{author}{\bibfnamefont{B.~I.} \bibnamefont{Halperin}},
  \bibinfo{author}{\bibfnamefont{P.~A.} \bibnamefont{Lee}}, \bibnamefont{and}
  \bibinfo{author}{\bibfnamefont{W.~F.} \bibnamefont{Brinkman}},
  \bibinfo{journal}{Phys.\ Rev. B} \textbf{\bibinfo{volume}{25}},
  \bibinfo{pages}{349} (\bibinfo{year}{1982}).

\bibitem[{\citenamefont{Schulz et~al.}(1982)\citenamefont{Schulz, Halperin, and
  Henley}}]{Schulz82}
\bibinfo{author}{\bibfnamefont{H.~J.} \bibnamefont{Schulz}},
  \bibinfo{author}{\bibfnamefont{B.~I.} \bibnamefont{Halperin}},
  \bibnamefont{and} \bibinfo{author}{\bibfnamefont{C.~L.}
  \bibnamefont{Henley}}, \bibinfo{journal}{Phys.\ Rev. B}
  \textbf{\bibinfo{volume}{26}}, \bibinfo{pages}{3797} (\bibinfo{year}{1982}).

\bibitem[{\citenamefont{Katifori and Nelson}(2006)}]{Katifori06}
\bibinfo{author}{\bibfnamefont{E.}~\bibnamefont{Katifori}} \bibnamefont{and}
  \bibinfo{author}{\bibfnamefont{D.~R.} \bibnamefont{Nelson}},
  \bibinfo{journal}{Phys.\ Rev. B} \textbf{\bibinfo{volume}{73}},
  \bibinfo{pages}{214503} (\bibinfo{year}{2006}).

\bibitem[{\citenamefont{Hatano and Nelson}(1997)}]{Hatano97}
\bibinfo{author}{\bibfnamefont{N.}~\bibnamefont{Hatano}} \bibnamefont{and}
  \bibinfo{author}{\bibfnamefont{D.~R.} \bibnamefont{Nelson}},
  \bibinfo{journal}{Phys.\ Rev. B} \textbf{\bibinfo{volume}{56}},
  \bibinfo{pages}{8651} (\bibinfo{year}{1997}).

\bibitem[{\citenamefont{Hatano and Nelson}(1998)}]{Hatano98}
\bibinfo{author}{\bibfnamefont{N.}~\bibnamefont{Hatano}} \bibnamefont{and}
  \bibinfo{author}{\bibfnamefont{D.~R.} \bibnamefont{Nelson}},
  \bibinfo{journal}{Phys. Rev. B} \textbf{\bibinfo{volume}{58}},
  \bibinfo{pages}{8384} (\bibinfo{year}{1998}).

\bibitem[{\citenamefont{Sow et~al.}(1998)\citenamefont{Sow, Harada, Tonomura,
  Crabtree, and Grier}}]{Chorng98}
\bibinfo{author}{\bibfnamefont{C.-H.} \bibnamefont{Sow}},
  \bibinfo{author}{\bibfnamefont{K.}~\bibnamefont{Harada}},
  \bibinfo{author}{\bibfnamefont{A.}~\bibnamefont{Tonomura}},
  \bibinfo{author}{\bibfnamefont{G.}~\bibnamefont{Crabtree}}, \bibnamefont{and}
  \bibinfo{author}{\bibfnamefont{D.~G.} \bibnamefont{Grier}},
  \bibinfo{journal}{Phys. Rev. Lett.} \textbf{\bibinfo{volume}{80}},
  \bibinfo{pages}{2693} (\bibinfo{year}{1998}).

\bibitem[{\citenamefont{Wadas et~al.}(1992)\citenamefont{Wadas, Fritz, Hug, and
  Güntherodt}}]{Wadas92}
\bibinfo{author}{\bibfnamefont{A.}~\bibnamefont{Wadas}},
  \bibinfo{author}{\bibfnamefont{O.}~\bibnamefont{Fritz}},
  \bibinfo{author}{\bibfnamefont{H.~J.} \bibnamefont{Hug}}, \bibnamefont{and}
  \bibinfo{author}{\bibfnamefont{H.~J.} \bibnamefont{Güntherodt}},
  \bibinfo{journal}{Z. Phys. B} \textbf{\bibinfo{volume}{88}},
  \bibinfo{pages}{317} (\bibinfo{year}{1992}).

\bibitem[{\citenamefont{Volodin et~al.}(1998)\citenamefont{Volodin, Temst,
  Haesendonck, and Bruynseraede}}]{Volodin98}
\bibinfo{author}{\bibfnamefont{A.}~\bibnamefont{Volodin}},
  \bibinfo{author}{\bibfnamefont{K.}~\bibnamefont{Temst}},
  \bibinfo{author}{\bibfnamefont{C.~V.} \bibnamefont{Haesendonck}},
  \bibnamefont{and}
  \bibinfo{author}{\bibfnamefont{Y.}~\bibnamefont{Bruynseraede}},
  \bibinfo{journal}{Appl. Phys. Lett.} \textbf{\bibinfo{volume}{73}},
  \bibinfo{pages}{1134} (\bibinfo{year}{1998}).

\end{thebibliography}

\end{document}